# Short Term Power Demand Prediction Using Stochastic Gradient Boosting

Ali Bou Nassif
Department of Electrical and Computer Engineering
University of Sharjah
Sharjah, United Arab Emirates
anassif@sharjah.ac.ae

*Abstract*— Power prediction demand is vital in power system and delivery engineering fields. By efficiently predicting the power demand, we can forecast the total energy to be consumed in a certain city or district. Thus, exact resources required to produce the demand power can be allocated. In this paper, a Stochastic Gradient Boosting (aka Treeboost) model is used to predict the short term power demand for the Emirate of Sharjah in the United Arab Emirates (UAE). Results show that the proposed model gives promising results in comparison to the model used by Sharjah Electricity and Water Authority (SEWA).

## I. INTRODUCTION

In 2003, a major power outage occurred in North America and was known as "Northeast blackout of 2003" [1]. During this blackout, 45 million people in USA and 10 million people in Ontario Canada were affected. The losses incurred by this outage exceeded 6 billion USD. In addition to the electric power outage, many sectors were affected. For instance, in some areas there were problems in water supply due to the deficiency in water pump pressure. Moreover, railroad services were interrupted , and there was a disruption in the cellular networks.

One of the main reasons of the Northeast blackout is the improper prediction of the power demand [1]. There are two types of improper prediction (1) underestimation and (2) overestimation. Underestimation occurs when electric engineers underestimate the actual demand load at a certain hour or in peak time. An example of underestimation can be seen in the following scenario:

In power engineering, the accelerating power must be zero under normal conditions as seen in the equation below [2].

*AcceleratingPower = PowerGenerated – PowerLoad*

When the Power Load increases, the Power Generated must increase accordingly to maintain the value of Accelerating Power to zero. If there are no sufficient resources to increase the Power Generated, the Accelerating Power will be negative. In this case, some of the load will be removed to balance the generated load and this will cause partial power outage. If the load remains the same, the speed of the generator will decrease and will cause the frequency to decrease. If the frequency decreases below a certain threshold (e.g. 49.50 Hz in Sharjah Emirate) and no loads will be removed, the main generator will shut down and this will lead to a power outage.

On the other hand, overestimation in power demand prediction will result in wasting some of the resources and will increase the cost of generating power.

There are three main types of load forecasting; (1) short term load forecasting, (2) medium term load forecasting and (3) long term load forecasting. The short term forecasting is commonly used to predict the load of the next hour, or even after 30 minutes but it can also be used to predict the power up to one week. Generators dispatching, voltage regulating, unit commitment, real-time pricing in the energy market are all examples of activities done during the short term prediction. On the contrary, medium term forecasting concerns forecasting of power demand for the next month. Long term forecasting deals with predicting the load for the next several months or years.

This paper is focused on the short term load forecasting using a Stochastic Gradient Boosting (aka Treeboost) model for the Emirate of Sharjah. The dataset used is provided by Sharjah Electricity and Water Authority (SEWA). The results of the proposed model are very promising as the mean relative error of the Treeboost model is 4% in comparison to 8% error in the model that SEWA is using.

The present paper is structured as follows: Section II gives a background about the work. Section III presents the related work where Section IV explains the methodology used. Finally, Section V presents the obtained results and provides some discussion about the results.

## II. BACKGROUND

### A. Power Load Forecasting

There are several factors that affect the power load forecasting. Examples include [3]:
1- Historical power load readings
2- Weather conditions (mainly temperature, humidity)
3- Customers' classes (commercial vs residential)
4- Time factor (hour / day / eek/ month)
5- Economic indicators (energy prices)

For short Time forecasting, in addition to historical load, weather conditions and time factor are very important. The time factor takes into consideration the exact time required to predict a load (e.g. 09:00 am on March 15). Moreover, the prediction of the load in the weekend is different from the load in week days. Weather conditions such as temperature and humidity play a pivotal role in the load forecasting. Figure 1 shows the average actual load (in Mega Watt) in 2014 of the Sharjah Emirate, UAE per month (1 = January, 2 = February, etc.). It is obvious that average load in July (month 7) is about 2.5 times the average load in January.

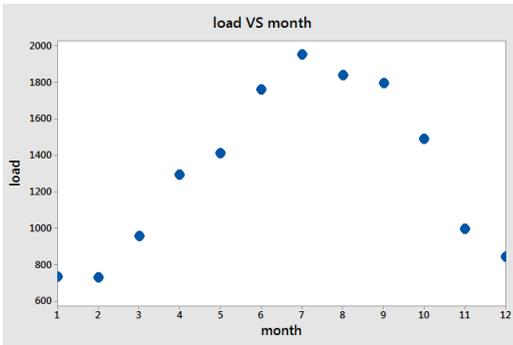

Figure 1: Average load forecast per month

The variation of humidity is also very important as shown in Figure 2. As seen in Figure 2, there is no well-known distribution for the humidity in Sharjah Emirate per month. The humidity might have large variations in the same month so it is vital to consider both temperature and humidity in the short term forecasting.

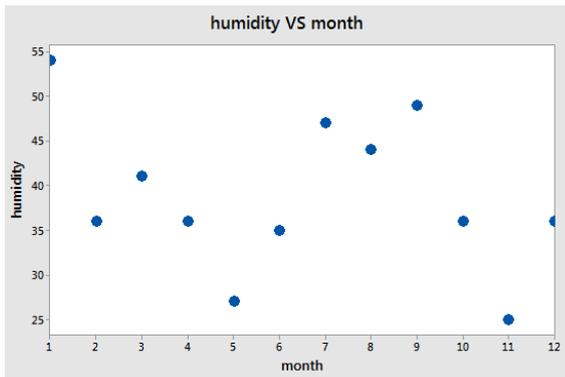

Figure 2: Humidity per month

*B. Stochastic Gradient Boosting*

Machine Learning methods based on gradient algorithms have been used in literature [4][5][6][7][8][9]. Stochastic Gradient Boosting (SGB) is also called Treeboost [10]. Boosting is a process to improve the accuracy of a predictive function by applying the function frequently in a series and combining the output of each function. The Treeboost model consists of a series of decision trees so the accuracy of the Treeboost is higher than a single decision tree. The main disadvantage of the Treeboost is that it is more complicated than a single decision tree and needs more computational resources. Moreover, the Treeboost model acts like a black box and cannot represent a big picture of the problem as a single decision tree does. The main characteristics of the Treeboost model are:

• The Treeboost model implements the regression part uses Huber-M loss function [11]. This function is a mixed of Least Absolute Deviation (LAD) and ordinary least squares (OLS) and. For residuals which are less than a cutoff point (Huber's Quantile Cutoff), the square of the residuals is used. Otherwise, absolute values are used. This method is used to overcome the problems raised from outliers. For outliers, squaring the residuals will lead to huge values, so they will be treated with the "absolute values" method instead. The Huber's Quantile Cutoff value is recommended to be between 0.9 and 0.95.

• In the Stochastic Gradient Boosting algorithm, "Stochastic" means that a random percentage of training data points (50% is recommended) will be used for each iteration instead of using all data for training. This yields to improvement in the results.

• The Stochastic Gradient Boosting (SGB) algorithm has a factor called Shrinkage factor. If each tree in the series is multiplied by this factor (between 0 and 1), it will delay the learning process and consequently, the length of the series will be longer to compensate for the shrinkage. This also leads to better prediction values.

The Treeboost algorithm is described as:

$$F(x) = F_0 + A1*T1(x) + A2*T2(x) + ... + AM*TM(x).$$

Where $F(x)$ is the predicted target, $F_0$ is the starting value, x is a vector which represents the pseudo-residuals, $T1(x)$ is the first tree of the series that fits the pseudo-residuals and A1, A2, etc. are coefficients of the tree nodes.

### III. RELATED WORK

Short term load forecasting using Machine Learning has been addressed in the past few decades. Examples of related work include:

P. Qingle and Z. Min [12] used neural network and rough set in short term load forecasting to improve the accuracy of model. Without rough set, the error of forecasting would be large so the authors used neural network and rough set in order to minimize the error and obtain high accuracy.

S. Ramos et al. [13] worked to develop short-term load forecasting by using holt-winter exponential smoothing and artificial neural network, at the end they compared between using holt-winter exponential smoothing and neural network.

W. Charytoniuk et al. [14] used nonparametric regression for short term load forecasting. This method is based on probability density function of load and factors that affecting to load. The accuracy of this model depends on the implemented historical data.

D. K. Ranaweera et al. [15] developed model base on fuzzy logic combined with historical weather load and load data for short term load forecasting, the results that obtained from fuzzy logic have similar accuracy compared to complicated statistical and back-propagation neural network.

S. Chenthur Pandian et al. used fuzzy logic in short term load forecasting. They considered time and temperature of day as input and load forecasting as output, the first variable (time) has eight triangular membership, the second variable (temperature) has four triangular membership, the load forecasting as output has eight triangular membership, at the end they compared the results of the output with conventional forecasted values and shows slightly math actual values.

A. M. Al-Kandari et al. [16] developed short term load forecasting model by using fuzzy liner regression. First, they converted the estimation fuzzy problem to linear optimization problem, then they tried to build program that is the base of simplex method. They concluded that by using this model they can obtain more reliable power system.

N. Amjady and F. Keynia [17] developed short-term load forecasting by using combination of wavelet transform (WT), neural network (NN) and evolutionary algorithm (EA), in order to obtain more accurate model.

Z. Xiao et al. [18] presented back propagation neural network combination with rough set for intricate short term load forecasting with dynamic and nonlinear factors to improve accuracy of forecasting, at the end, the model was tested and gave good performance compared to the back propagation.

There are two main contributions of our work in comparison to previous work. First, to the best of our knowledge, this is the first work that addresses the short term load prediction using Stochastic Gradient Boosting. Secondly, this is the first work that predicts the short term power demand for Sharjah Emirate. Since the weather in Sharjah is very similar to many cities in the Gulf Countries, the proposed model can be easily modified to predict the short term load for casting to these countries as well.

## IV. METHODOLOGY

### A. Dataset

The dataset used in this research was collected from Sharjah Electricity and Water Authority (SEWA) and the Meteorological Office of Sharjah International Airport (MET Office). We collected data for years 2014 and 2015. The data contains the actual load (in Mega Watt) for the Sharjah Emirate taken every 30 minutes. The temperature and the humidity were taken from the MET office and the dataset was cleaned so that it contains the actual load in addition to temperature and humidity every 30 minutes.

Table 1 depicts some statistics regarding the actual load, temperature and humidity in 2014.

Table 1: Descriptive statistics for model variables

| Variable | Mean | StDev | Min | Median | Max | Skew | Kurtosis |
|---|---|---|---|---|---|---|---|
| load | 1242 | 432.3 | 481 | 1270 | 2083 | 0.05 | -1.42 |
| temp | 28.339 | 7.49 | 10.7 | 29 | 53.4 | -0.14 | -0.87 |
| humidity | 53.995 | 18.512 | 4 | 54 | 98 | -0.07 | -0.69 |

As noticed from Table 1, there is a noticeable variation in the three variables between the minimum and maximum values. In order to study the type of distribution, the histogram of each variable is displayed in Figures 3, 4 and 5.

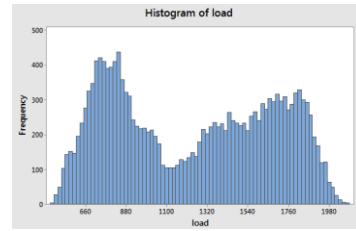
Figure 3: Histogram of load

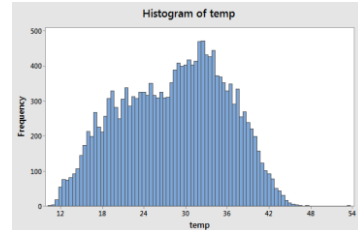
Figure 4: Histogram of temperature

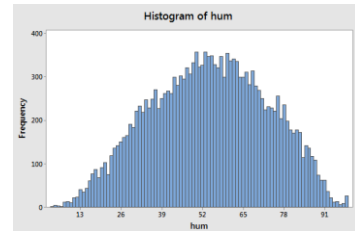
Figure 5: Histogram of humidity

Based on the above figures, one can notice that the humidity variable is normally distributed. The distribution of the temperature is slightly close to normal with negative skew. On the other hand, the load is not normally distributed. So, it is not possible to take a sample data randomly to predict the load at a certain time because this load varies every 30 minutes. In our research, we followed the following methodology: in order to predict the load at a certain time $t_0$, we wrote a MATLAB script to extract the dataset required to train the Treeboost model. For the extracted dataset, we choose a time frame of 3 hours prior to time $t_0$ (these are 6 data points because we have a data point for each 30 minutes). Then we take the same time frame for last 10 days excluding Friday. So, the size of the extracted dataset will be 60 data points. We excluded Fridays because Fridays are holidays and the load in these days is much less than the load of the other days. In our experiments, we selected 12 datasets; each belongs to a different month.

### B. Training and Testing The Treeboost Model:

The parameters of the proposed model are as follows:
- Maximum number of trees: 400
- Depth of individual tree: 5
- Minimum size node to split: 10
- Huber's quantile cutoff: 0.95
- Influence trimming factor: 0.01

As explained before, the extracted dataset is 60 rows. The upper row contains the time $t_0$ where its load is predicted. So,

the upper row (row #1) will be used for testing the model and the remaining 59 rows will be used to train the model.

### C. SEWA Methodology for predicting the load

While visiting SEWA in Sharjah, the concerned engineer explained the methodology used by SEWA to predict the load for a day which can be considered as a linear regression. The engineer monitors the load every day in the morning from 7 am until 10 am. Then he compares these readings with the readings of the previous working day for the same time frame. The maximum difference (the difference can be positive or negative) in load between the day and its previous working day will be added to the load of the previous day to predict the load of the current day after 10 am.

### D. Evaluation Criteria

The prediction accuracy of the models is assessed using the Mean Absolute Error (MAE) and the Mean Magnitude of Relative Error (MMRE) criteria.

*MAE* computes the difference in absolute value between actual ($x_i$) and predicted ($\hat{x}_i$) load values.

*MMRE* computes the mean of the absolute percentage of error between actual ($x_i$) and predicted ($\hat{x}_i$) load values as shown in the equations below:

$$MAE_i = |x_i - \hat{x}_i| \quad MRE_i = \frac{|x_i - \hat{x}_i|}{x_i} \quad MMRE = \frac{1}{N}\sum_{i=1}^{N} MRE_i$$

## V. RESULTS AND DISCUSSION

This section displays and discusses the results. Figure 6 shows the scatterplot diagram for the proposed SEWA model versus the SEWA model.

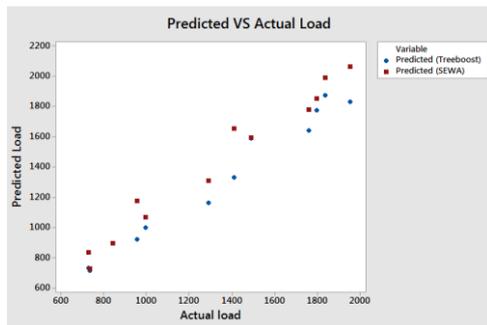

Figure 6: Treeboost VS SEWA models

Figure 6 shows that the predicted load in the Treeboost model is closer to the actual load than of SEWA's model and thus the Treeboost model outperforms the SEWA model. To confirm our conclusions, we calculated the MAE and MMRE as shown in Table 2:

Table 2: Comparison between Treeboost and SEWA

| Model / Criteria | Treeboost | SEWA |
|---|---|---|
| MAE (Mega Watt) | 60 | 97 |
| MMRE (%) | 4.3 | 8 |

Table 2 demonstrates that the Treeboost model surpasses the SEWA model because it gives less error and can be used as an alternative for short term load forecasting.

Future work will focus on developing other Machine Learning models for short term forecasting.

## Acknowledgement:

Dr. Ali Bou Nassif would like to thank the University of Sharjah, Sharjah, UAE for supporting and funding this work though the Seed Grant number: 1602040221 – P. We would also like to thank SEWA for providing the dataset.